\documentclass[12pt,twocolumn]{aastex63}

\usepackage{lmodern}
\usepackage[T1]{fontenc}
\usepackage[nice]{nicefrac}
\usepackage{amsmath}
\usepackage{threeparttable}
\usepackage{longtable}
\usepackage{threeparttablex}

\newcommand{\swift}{\textit{Swift}}
\newcommand{\fermi}{\textit{Fermi}}
\usepackage{lineno}

\received{August 25, 2020}
\revised{October 16, 2020}
\accepted{October 18, 2020}
\published{\today}
\submitjournal{ApJ}

\shorttitle{160625B}
\shortauthors{Cunningham et al.}

\begin{document}

\title{GRB\,160625B: Evidence for a Gaussian-Shaped Jet}

\correspondingauthor{Virginia Cunningham}
\email{vcunning@astro.umd.edu}

\author[0000-0003-2292-0441]{Virginia Cunningham}
\affiliation{Department of Astronomy, University of Maryland, College Park, MD 20742, USA}
\affiliation{Center for Research and Exploration in Space Science and Technology, NASA/GSFC, Greenbelt, MD 20771}

\author[0000-0003-1673-970X]{S. Bradley Cenko}
\affiliation{Astrophysics Science Division, NASA Goddard Space Flight Center, MC 661, Greenbelt, MD 20771, USA}
\affiliation{Joint Space-Science Institute, University of Maryland, College Park, MD 20742, USA}
\author[0000-0001-9068-7157
]{Geoffrey Ryan}
\affiliation{Joint Space-Science Institute, University of Maryland, College Park, MD 20742, USA}
\author{Stuart N. Vogel}
\affiliation{Department of Astronomy, University of Maryland, College Park, MD 20742, USA}
\author{Alessandra Corsi}
\affiliation{Department of Physics and Astronomy, Texas Tech University, Box 1051, Lubbock, TX 79409-1051, USA}
\author{Antonino Cucchiara}
\affiliation{College of Marin/University of the Virgin Islands, 120 Kent Ave., Kentfield, CA 94904, USA}
\author[0000-0002-6652-9279]{Andrew S. Fruchter}
\affiliation{Space Telescope Science Institute, 3700 San Martin Drive, Baltimore, MD 21218, USA}
\author{Assaf Horesh}
\affiliation{Racah Institute of Physics, The Hebrew University of Jerusalem, Jerusalem, 91904, Israel}
\author[0000-0002-5477-0217]{Tuomas Kangas}
\affiliation{Space Telescope Science Institute, 3700 San Martin Drive, Baltimore, MD 21218, USA}
\author{Daniel Kocevski}
\affiliation{Astrophysics Office, ST12, NASA/Marshall Space Flight Center, Huntsville, AL 35812, USA }
\author{Daniel A. Perley}
\affiliation{Astrophysics Research
Institute, Liverpool John Moores University, Liverpool Science Park, 146 Brownlow Hill, Liverpool L35RF, UK}
\author{Judith Racusin}
\affiliation{Astrophysics Science Division, NASA Goddard Space Flight Center, MC 661, Greenbelt, MD 20771, USA}

\begin{abstract}

We present multiwavelength modeling of the afterglow from the long $\gamma$-ray burst GRB\,160625B using Markov Chain Monte Carlo (MCMC) techniques of the \texttt{afterglowpy} Python package. GRB\,160625B is an extremely bright burst with a rich set of observations spanning from radio to $\gamma$-ray frequencies. These observations range from $\sim$0.1 days to $>$1000 days, thus making this event extremely well-suited to such modeling. In this work we compare top-hat and Gaussian jet structure types in order to find best fit values for the GRB jet collimation angle, viewing angle, and other physical parameters. We find that a Gaussian-shaped jet is preferred (2.7-5.3$\sigma$) over the traditional top-hat model. Our estimate for the opening angle of the burst ranges from 1.26$^{\circ}$ to 3.90$^{\circ}$, depending on jet shape model. We also discuss the implications that assumptions on jet shape, viewing angle, and particularly the participation fraction of electrons have on the final estimation of GRB intrinsic energy release and the resulting energy budget of the relativistic outflow. Most notably, allowing the participation fraction to vary results in an estimated total relativistic energy of $\sim10^{53}$ erg. This is two orders of magnitude higher than when the total fraction is assumed to be unity, thus this parameter has strong relevance for placing constraints on long GRB central engines, details of the circumburst media, and host environment. 

\end{abstract}


\section{Introduction} \label{sec:intro}

Long $\gamma$-ray bursts (GRBs)\footnote{The primary focus throughout this paper will be on long-duration GRBs, unless otherwise noted.} are amongst the most violent and energetic phenomena in the Universe. Despite observations of thousands of GRBs over the last few decades, key open questions -- such as the nature of the central engine and the structure and composition of the relativistic jets -- remain unsolved.

One key to unraveling these mysteries lies in accurately measuring their energetics. Estimates of the total relativistic energy released by a GRB can have major implications for constraining their physical characteristics and origins. Precise measurements could potentially distinguish between different progenitor systems. Two popular theories include rotationally-powered magnetars \citep{Zhang2001,Thompson2004,Metzger2015} and the collapse of a massive star into a black hole \citep{Woosley1993,Macfadyen1999,Woosley2012}.

GRBs are known to be highly collimated explosions with jet opening angles typically between 1--10$^{\circ}$ \citep{Sari1999,Rhoads1999}. The true value of the intrinsic energy release of a GRB is dependent upon this collimation angle: $E_{\gamma} = E_{\gamma,\mathrm{iso}}(1-\mathrm{cos} \ \theta_{j}) \approx E_{\gamma,\mathrm{iso}} \frac{\theta_{j}^{2}}{2}$, where $E_{\gamma}$ is the beaming-corrected $\gamma$-ray energy of the burst, $E_{\gamma,\mathrm{iso}}$ is the uncorrected isotropic $\gamma$-ray energy, and $\theta_j$ is the jet half-opening angle\footnote{This equation is only valid for the simple top-hat jet. More complicated jet structure types are discussed in \S\ref{sec:Gaussian}} \citep{Bloom2001,Frail2001}. This jet collimation correction can affect the value of $E_{\gamma}$ by a factor of 10 -- 100 \citep{Frail2001}. Therefore a precise measurement of $\theta_j$ is imperative for understanding the true energetics of GRBs.

Making a precise measurement of the collimation angle can be difficult however, as it usually requires sustained, detailed, multiwavelength observations of the GRB afterglow and the identification of a `jet-break', i.e., a change in the temporal slope of the light curve associated with the observer becoming aware of the edge of the jet \citep{Sari1999,Panaitescu2007,Kocevski2008,Racusin2009,Goldstein2016}. Alternatively, the energy of the explosion can be inferred via non-relativistic calorimetry \citep{Frail2000,Berger2004}. At late times the ejecta slows to a non-relativistic spherical blastwave and can be modeled independently of the jet collimation angle. This is of course only possible when sufficiently late-time radio data exists. 

\begin{figure*}
\includegraphics[trim=10 10 60 50,clip,width=1.0\linewidth]{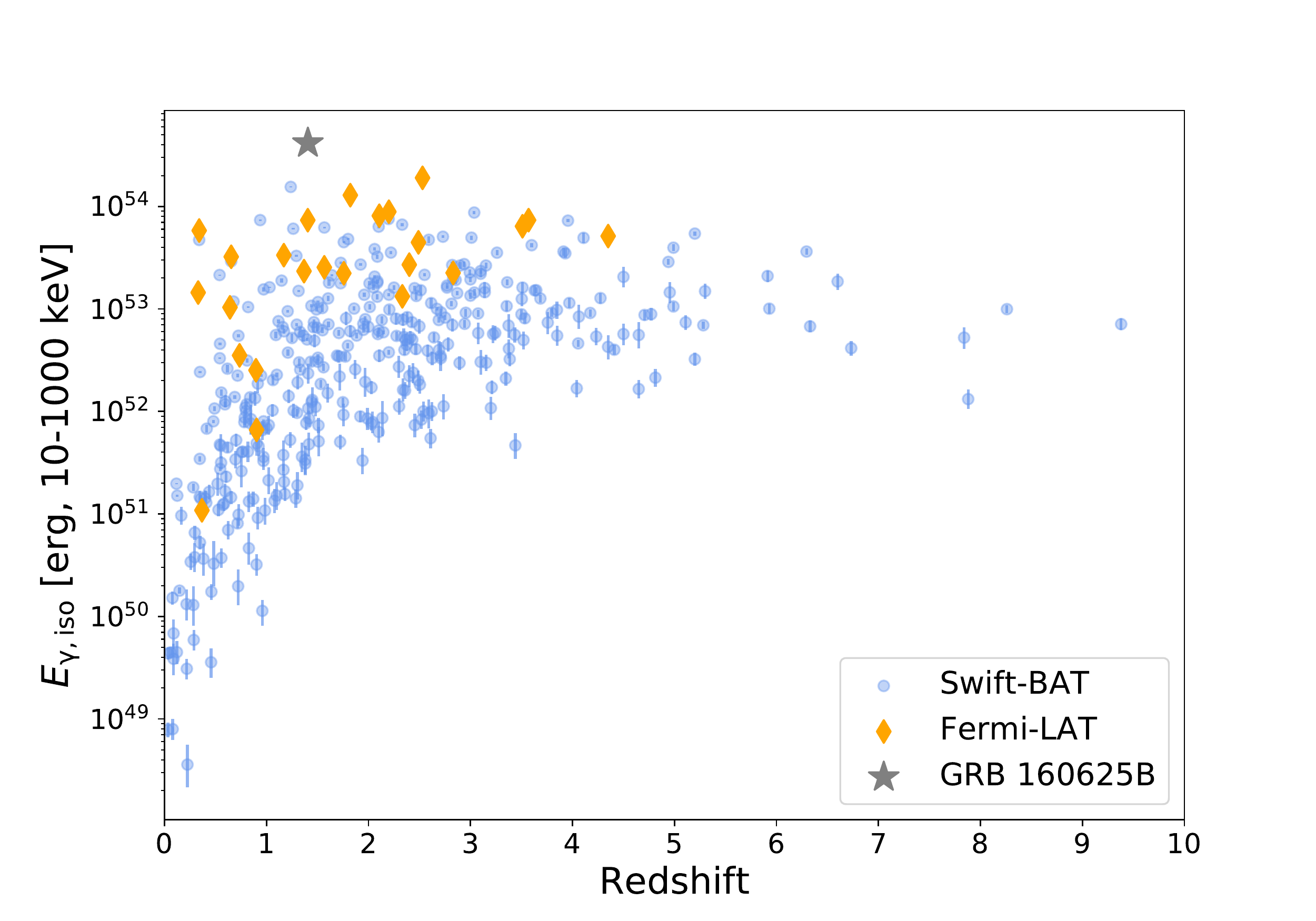}
\caption{Bolometric (10-1000 keV) isotropic energy release vs redshift compared between LAT-detected and BAT-detected GRBs. In general, GRBs detected by LAT tend to be brighter, more energetic events and GRB\,160625B is itself extreme even for this population. Data are taken from \citet{Lien2016} and \citet{Ajello2019}. 
\label{fig:eiso_z}}
\end{figure*}

Here we focus on events detected at GeV energies by the Large Area Telescope (LAT; \citealt{Atwood2009}) on \fermi. This sample is well-suited for studying GRB energetics since \fermi\ tends to select events with high values of $E_{\mathrm{\gamma,iso}}$. This effect can be partly explained by both the Amati Relation ($E_{peak} - E_{\gamma,\mathrm{iso}}$, \citealt{Amati2009}) and also the lower sensitivity of the LAT compared to X-ray instruments. LAT-detected GRBs often display values of $E_{\mathrm{\gamma,iso}} > 10^{53}$ erg (Figure \ref{fig:eiso_z}; \citealt{Cenko2010,Cenko2011,Xu2013,Perley2014}) and their afterglows can generally be well modeled by a series of more simple power laws \citep{Yamazaki2019}.

We are conducting a campaign to model the broadband behavior of a sample of LAT-detected GRBs. Here we present the methodology and apply this to one example, that of GRB\,160625B - an exceptionally bright long-GRB at z=1.406 \citep{Xu2016}. Future work will discuss the broader population in the hopes of identifying whether LAT GRBs in fact represent a unique group and if so, how they differ from the rest of the GRB population (e.g., jet shape, local environments, magnetar vs black hole central engines; \citealt{Racusin2011}). 

Several previous works have already performed detailed afterglow analysis of GRB\,160625B (\citealt{Troja2017,Alexander2017,Kangas2020}, hereafter T17, A17 and K20, respectively). We expand on these works by combining and including all available data as well as undertaking new late-time observations with the Very Large Array (VLA). We also expand the analysis by considering additional models for the internal structure of the GRB jet beyond the canonical top-hat model as well as taking into account viewing angle effects, quantifying the uncertainty of derived parameters, and investigating the role individual afterglow parameters, such as the participation fraction, $\xi$, have on the model fitting. 

For our modeling of GRB\,160625B, we make use of the new \texttt{afterglowpy} software package \citep{Ryan2020}. \texttt{Afterglowpy} is a publicly available open-source Python package for the numerical computation of structured jet afterglows. This package is unique in its ability to test a range of jet structures such as top-hat, Gaussian, and power law while leaving the viewing angle and jet collimation angle as free parameters. There is evidence to suggest that the interaction between the GRB jet and the surrounding medium could take on a variety of forms - based on numerical simulations \citep{Alloy2005, Duffell2013b, Margutti2018} and observations of GW\,1701817/GRB\,170817A \citep{Lazzati2018,Troja2018}. In fact, most GRB jets probably deviate significantly from the simple on-axis top-hat model (e.g., \citealt{Ryan2015}) and \citet{Strausbaugh2019} has already suggested GRB\,160625B can be modeled as a structured jet so it is imperative to consider this when modeling GRB afterglows. 

This paper is organized as follows: we describe the available data products and our data reduction methods in \S \ref{sec:data}. In \S \ref{sec:model} we define the details of the GRB afterglow modeling. We discuss the implications of our results in \S \ref{sec:discussion} and summarize our conclusions in \S \ref{sec:conclusions}. All error bars correspond to $1\sigma$ uncertainties and we assume a standard $\Lambda$CDM cosmology \citep{Planck2018} throughout the analysis. 

\section{Observations and Data Reduction} \label{sec:data}

\subsection{\texorpdfstring{$\gamma$-ray}{Gamma-ray}}
The \fermi\ Gamma-ray Burst Monitor \citep[GBM;][]{Meegan2009} first triggered on GRB\,160625B at 22:40:16.28 UT on 25 June 2016 (UT times are used throughout this work) and again at 22:51:16.03. The burst is characterized by three `sub pulses' separated by two relatively quiescent periods over a duration of $t_{90}=460$s (50-300 keV) and the spectral shape of the burst is well modeled by a Band function  \citep{Burns2016}. The fluence in the 10-1000 keV bandpass was $(6.4256 \pm 0.0019) \times 10^{-4}$ erg cm$^{-2}$ \citep{vonKienlin2020}. 

The \fermi\ Large Area Telescope \citep[LAT;][]{Atwood2009} triggered on the second pulse at 22:43:24.82. The GRB location was in the LAT field-of-view for $\sim$1000 seconds after its initial trigger (20 MeV to 300 GeV, \citealt{Dirirsa2016}). The highest energy photon observed in the rest-frame was 15.3 GeV which occurred $\sim$345 seconds after the first GBM trigger \citep{Ajello2019}. 

The GRB was also detected by Konus-\textit{Wind} ($9.44\pm0.16\times10^{-4}$ erg cm$^2$) from 20 keV to 10 MeV, \citealt{Svinkin2016}), SPI-ACS/INTEGRAL \citep{Kann2016}, and CALET \citep{Nakahira2016}. For the anaylsis presented here we choose $t_0$ to be that corresponding to the first GBM trigger\footnote{Most other analyses reference $t_0$ to that of the LAT trigger, but since we are ignoring very early data we consider any differences negligible.}. 

\subsection{X-Ray}
The \textit{Neil Gehrels Swift Observatory} \citep{Gehrels2004} began observing GRB\,160625B 2.5 hours after the initial GBM trigger \citep{Evans2016}. The X-ray Telescope (XRT; \citealt{Burrows2005}) on board \swift\ observed GRB\,160625B for 47 days. XRT data are taken from the publicly available online \swift\ burst analyzer tool\footnote{https://www.swift.ac.uk/burst\_analyser/00020667/}. For details of how these light curves were produced, see \citet{Evans2007,Evans2009}. The hardness ratio appears relatively constant over time so we assume a single spectrum that can be described as an absorbed power law with the Galactic neutral hydrogen column fixed to $9.76 \times 10^{20}$ cm$^{-2}$ \citep{Willingale2013}. Using a photon index of $\Gamma_x$= 1.86$^{+0.10}_{-0.09}$, assuming an intrinsic host absorption of $n_{\mathrm{H,int}} = 1.8 \times 10^{21}$ cm$^{-2}$, an unabsorbed counts-to-flux conversion factor of $4.4 \times 10^{-11}$ erg cm$^{-2}$ ct$^{-1}$, and a redshift of 1.406 \citep{Xu2016} we convert the 0.3-10 keV flux light curves to a flux density at an energy of 5 keV. We choose only to include data taken during photon counting mode, which begins 0.1 days after the burst. In addition to the XRT data we include the late-time \textit{Chandra} observations taken by K20 at 69.8 and 144 days after the burst. All X-ray observations are available in Table \ref{tab:xray}.

\begin{table}
  \centering
 \caption{X-ray Data }
  \begin{tabular}{cccccc}
    \toprule
$\Delta$t & Energy & Flux Density & Instrument \\

[day] & [keV] & [nJy] & \\
\hline
0.12 & 5 & 4787 $\pm$ 1076 & \textit{Swift}/XRT \\
0.12 & 5 & 4358 $\pm$ 981 & \textit{Swift}/XRT \\
0.12 & 5 & 5671 $\pm$ 1244 & \textit{Swift}/XRT \\
0.12 & 5 & 4772 $\pm$ 1073 & \textit{Swift}/XRT \\
 ... & ... & ... & ... \\
 41.31 & 5 & 1.34 $\pm$ 0.38 & \textit{Swift}/XRT \\
47.16 & 5 & 1.06 $\pm$ 0.47 & \textit{Swift}/XRT \\
69.76 & 5 & 0.61 $\pm$ 0.12 & \textit{Chandra}/ACIS-S \\
144.36 & 5 & 0.13 $\pm$ 0.03 & \textit{Chandra}/ACIS-S \\
\hline
  \end{tabular}
   \begin{tablenotes}
      \footnotesize
      \item  1. Times are in reference to the first GBM trigger (Jun 25 2016 22:40:16.28 UTC).
 \item 2. This table is available in its entirety in a machine-readable format. A portion is shown here for guidance regarding its form and content.
    \end{tablenotes}
  \label{tab:xray}
\end{table}


\subsection{Optical}
One defining feature of GRB\,160625B is its extremely bright optical afterglow and the presence of an optical `bump' around the time of the jet-break. \citet{Strausbaugh2019} suggest this excess emission could be the result of an edge-brightened jet while K20 suggest it could instead be produced by density fluctuations within the circumburst medium or angular brightness differences.

A17 utilized several optical instruments to observe the GRB -- the 2 m Faulkes Telescope North (FTN) operated by Las Cumbres Observatory (LCO), the  2 m Liverpool Telescope (LT) at Roque de los Muchachos Observatory (ORM), and the Low Dispersion Survey Spectrograph 3 (LDSS3) at Magellan -- ranging from 0.56 to 37 days post trigger. T17 observed the GRB with the Reionization And Transients InfraRed camera (RATIR) beginning 8 hours after the trigger until it faded beyond detection at $\sim$50 days and also reported u-band observations taken with the Ultraviolet/Optical Telescope (UVOT; \citealt{Roming2005}) on board \swift. In addition, several observations used in this work were compiled from the Gamma-ray Burst Coordinates Network (GCN) Circulars by \citet{Zhang2018} and appropriately converted to flux densities. Late time Hubble Space Telescope (HST) observations were reported by K20 71.5 and 140.2 days post trigger. The flux contribution from the host was already subtracted out by K20 and we account for Galactic extinction in the direction of the GRB, E(B-V) = 0.1107 mag \citep{Schlafly2011}, by assuming the extinction law described in \citet{Fitzpatrick1999}. All optical observations are available in Table \ref{tab:opt}.

%

\begin{table*}
  \centering
 \caption{Optical Data }
  \begin{tabular}{cccccc}
    \toprule
$\Delta$t & Filter & AB Mag & Frequency & Flux Density & Instrument \\

[day] & & & [$\times 10^{14}$ Hz] & [$\mu$Jy] & \\
\hline

     
0.37 &  r &  18.24 $\pm$ 0.01 & 4.82 & 240 $\pm$ 2 & RATIR \\
0.39 &  r &  18.29 $\pm$ 0.01 & 4.82 & 229 $\pm$ 2 & RATIR \\
0.41 &  r &  18.35 $\pm$ 0.01 & 4.82 & 216 $\pm$ 2 & RATIR \\
0.43 &  r &  18.43 $\pm$ 0.01 & 4.82 & 202 $\pm$ 2 & RATIR \\
 ... & ... & ... & ... & ... & ... \\
37.92 &  R &  23.68 $\pm$ 0.10 & 4.68 & 1.57 $\pm$ 0.15 & SAORAS/BTA \\
40.29 &  R &  23.52 $\pm$ 0.10 & 4.68 & 1.82 $\pm$ 0.18 & Maidanak/AZT-22 \\
44.34 &  R &  23.90 $\pm$ 0.11 & 4.68 & 1.28 $\pm$ 0.14 & Maidanak/AZT-22 \\
44.34 &  R &  <23.01 & 4.68 & <2.91 & Maidanak/AZT-22 \\ 
\hline
  \end{tabular}
   \begin{tablenotes}
      \footnotesize
      \item 1. Magnitudes are not corrected for extinction, while flux densities are. Times are in reference to the first GBM trigger (Jun 25 2016 22:40:16.28 UTC).
 \item 2. This table is available in its entirety in a machine-readable format. A portion is shown here for guidance regarding its form and content.
    \end{tablenotes}
  \label{tab:opt}
\end{table*}




\subsection{Radio}
We consolidate previous Karl G. Jansky VLA observations (Program IDs 15A-235 and S81171, PIs Berger and Cenko, respectively) from 1.37 to 209 days after the burst for the most complete sample of radio data (A17, K20). We obtained additional late time observations of GRB\,160625B taken at 6 GHz (C-band) on 4 Feb 2020 15:14:24 (1319 days post-burst; Program ID SC1031, PI Cenko) for an on-source integration time of 1.8 hours. The data were reduced using the standard VLA calibration pipeline provided by the Common Astronomy Software Applications package (CASA; \citealt{CASA}).  We use J2049+1003 as the complex gain calibrator and 3C286 as both the bandpass and flux calibrator. Imaging is done using the TCLEAN algorithm in CASA. We do not detect any emission at the afterglow location and so report a $3\sigma$ upper limit of $7.4\ \mu$Jy. We calculate this limit as three times the RMS uncertainty at the position of the GRB. In addition, we include data from the Australian Telescope Compact Array (ATCA) taken by T17 between 4.5 and 29 days post burst. All radio observations are available in Table \ref{tab:radio}.

\begin{table}
  \centering
 \caption{Radio Data }
  \begin{tabular}{cccccc}
    \toprule
$\Delta$t & Frequency & Flux Density & Instrument \\

[day] & [GHz] & [$\mu$Jy] & \\
\hline
1.37 & 5.00 & 163 $\pm$ 34 & VLA \\
1.37 & 7.10 & 232 $\pm$ 22 & VLA \\
1.35 & 8.50 & 288 $\pm$ 23 & VLA \\
1.35 & 11.00 & 507 $\pm$ 35 & VLA \\
 ... & ... & ... & ... \\
58.25 & 6.10 & 75 $\pm$ 10 & VLA \\
58.25 & 22.00 & 52 $\pm$ 12 & VLA \\
208.95 & 6.10 & 16 $\pm$ 5 & VLA \\
1319 & 6.10 & <2.46 & VLA \\
\hline
  \end{tabular}
   \begin{tablenotes}
      \footnotesize
      \item  1. Times are in reference to the first GBM trigger (Jun 25 2016 22:40:16.28 UTC).
 \item 2. This table is available in its entirety in a machine-readable format. A portion is shown here for guidance regarding its form and content.
    \end{tablenotes}
  \label{tab:radio}
\end{table}

\section{Afterglow Modeling \label{sec:model}}
\subsection{Basic Tenets \label{sec:tenets}}

The primary focus of this section is to model the broadband afterglow emission of GRB\,160625B with the goal of measuring the total energy output of the GRB central engine, the jet opening angle, and the geometry of the jet structure. Here, we assume the standard fireball model for the afterglow where the observed emission is synchrotron radiation from electrons in the circumburst medium accelerated by the relativistic blast wave \citep{Sari1998, Granot2002}. The emitting electrons are accelerated to a power-law distribution of energies with an index of $-p$ and a minimum Lorentz factor of $\gamma_m$. The resulting spectral energy distribution (SED) can be described by a series of power law segments smoothly broken at three characteristic frequencies -- $\nu_{sa}$, the self-absorption frequency, $\nu_{m}$, the frequency of the lowest energy electron in the distribution, and $\nu_c$, the cooling frequency. The values of the frequencies depend on the structure of the surrounding medium of the explosion as well as the jet shape, jet microphysics, energy produced, and viewing angle.

Before calculating detailed models, we infer the circumburst density profile via the temporal decline rate of the observations. From t $\sim$ 0.1 days to t = 20 days the optical and X-ray data can be well approximated by a single power law. Then at around 20 days the decay steepens, signifying the jet-break\footnote{\citet{Strausbaugh2019} place the jet-break slightly earlier during the peak of the optical bump at $\sim$13 days}. For the early time i-band data we find $\alpha_{\mathrm{i}} = -1.00 \pm 0.02$ and for the early XRT observations we find $\alpha_{\mathrm{X}} = -1.26 \pm 0.02$. 
The steepening of the slope between the optical and X-ray observations suggests that the cooling frequency, $\nu_c$, lies between these two regimes. In the case of a slow-cooling constant-density (ISM-like) profile $\alpha_{ISM} = \frac{3(1-p)}{4} \ (\nu < \nu_c)$ which yields an estimate for $p$, the index of the electron energy distribution, of $2.33\pm0.03$ \citep{Granot2002}. In the slow-cooling wind-like scenario $\alpha_{wind} = \frac{(1-3p)}{4}$ which yields $p = 1.67\pm0.03$. When $\nu > \nu_c$ the decline rate can be described by $\alpha = \frac{(2-3p)}{4}$ for both the wind and ISM-like profiles. Using $\alpha_{\mathrm{X}} = -1.26 \pm 0.02$ we find $p = 2.35\pm 0.03$, consistent with the ISM result. Despite fewer observations available at later times ($>20$ days) the optical decline rate of $\alpha_{\mathrm{i,late}} = -1.61 \pm 0.16$ produces $p = 2.15 \pm 0.21$ ($\alpha = \frac{3p}{4}$) and the X-ray decline rate of $\alpha_{\mathrm{X,late}} = -2.10 \pm 0.15$ produces $p = 2.46 \pm 0.20$ ($\alpha = \frac{3p-1}{4}$; \citealt{Ryan2020}). The early and late-time behavior are consistent; thus throughout this work we assume an ISM-like density profile for GRB\,160625B.\footnote{A17, T17, K20 and other works come to the same conclusion.} 

\begin{figure*}
\includegraphics[trim=50 10 80 40,clip,width=0.99\linewidth]{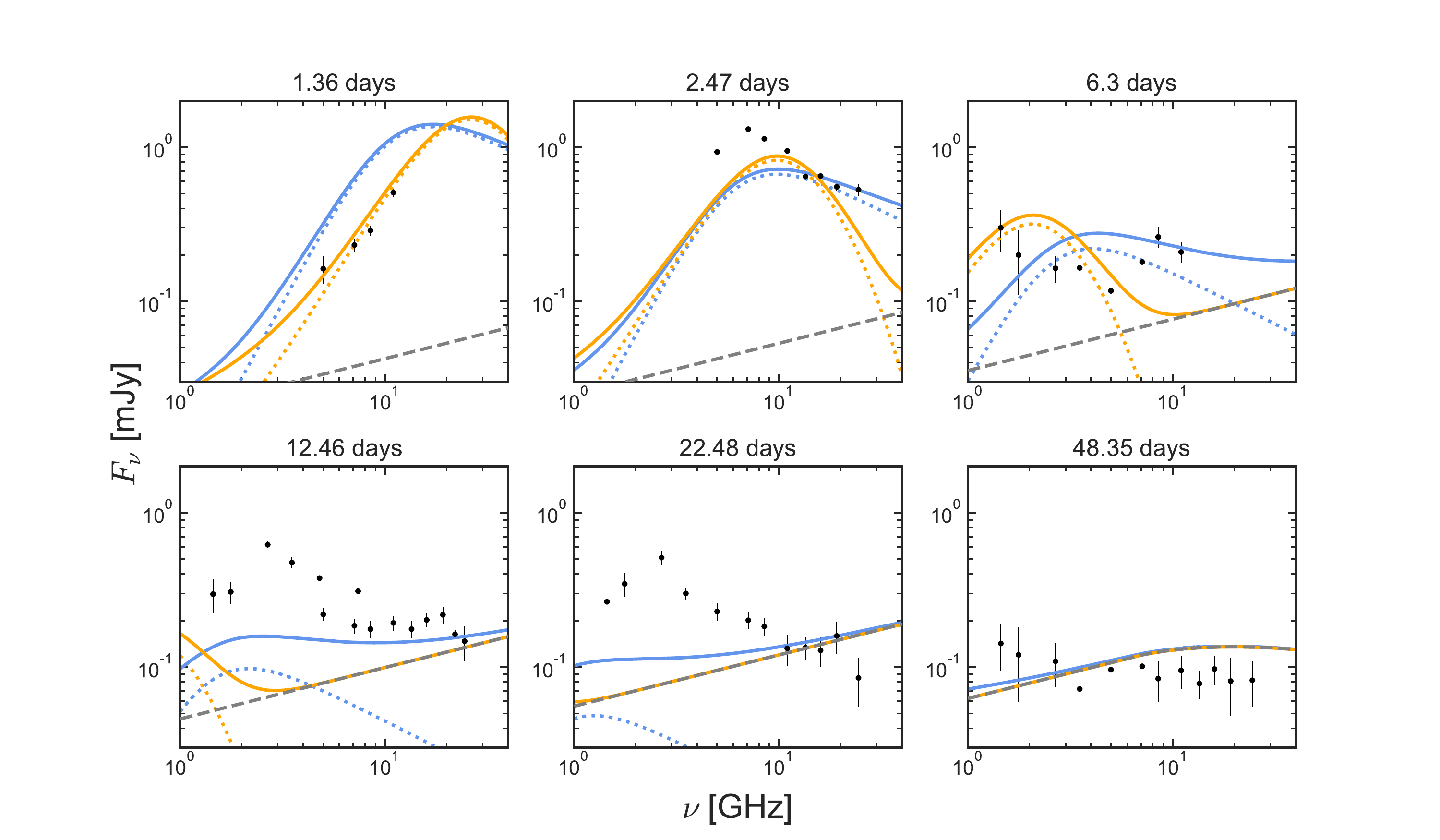}
\caption{The observed spectral energy distributions (SEDs) for the radio data at various epochs overplotted with forward and reverse shock models. Our forward shock model (gray dashed line) is as described in $\S$\ref{sec:xi}. The two reverse shock models (dotted lines) are taken from A17 and represent different assumptions for the location of the SED peak, $\nu_p$: $\nu_p = \nu_a$ (blue) and $\nu_p = \nu_c$ (orange). The solid lines represent the combination of both the forward and reverse shock models. Clearly the reverse shock, regardless of which model is used, dominates at early times ($<20$ days) and lower frequencies ($<10$ GHz).
\label{fig:rad_sed}}
\end{figure*}

There are significant features present in the early time radio data of GRB\,160625B which are likely not related to the forward shock emission (Figure \ref{fig:rad_sed}). A17 attribute these effects to the combination of a reverse shock and interstellar scintillation. Therefore when modeling the forward shock we conservatively choose to only include post-jet-break radio data ($\gtrsim20$ days) and at frequencies above 10 GHz (except for our late-time observation at 6.1 GHz) to mitigate these effects. A17 found there is negligible extinction due to dust within the GRB host galaxy and so we choose to ignore those effects in our analysis. Due to a systematic offset between the r- and i-band data of A17 and T17 we choose to only include that of T17 and none from A17 in our forward shock modeling as it is a larger dataset and observations were directly compared to the PanSTARRS magnitude system (see also K20).

\subsection{\texttt{Afterglowpy} Package \label{sec:afterglowpy}}
\texttt{Afterglowpy} uses the single-shell approximation to numerically and analytically model a blast wave propagating through an ISM-like circumburst medium as a function of viewing angle and jet type \citep{Ryan2020}. A range of jet structure types have been proposed in the literature -- some common examples include variations of top-hat, Gaussian, and power law models \citep{Meszaros1998, Rossi2002, Zhang2002, Rossi2004, Zhang2004, Alloy2005, Duffell2013b, Margutti2018, Coughlin2020}. The exact structure of any one GRB jet may be dependent upon several factors such as the immediate circumburst environment and interactions with the stellar envelope. Compared to other available afterglow modeling codes (e.g., \texttt{BoxFit}; \citealt{vanEerten2012}) \texttt{afterglowpy} is advantageous for its ability to probe this complex inner structure of the GRB jet. 

We employ the statistical sampling techniques of the EMCEE Python package for Markov-Chain Monte Carlo (MCMC; \citealt{Foreman-Mackey2013}) analysis with the \texttt{afterglowpy} models, as outlined in \citet{Troja2018}. \texttt{Afterglowpy} generates samples from the entire posterior distribution for each of the models we consider here in this work. As input, our fit takes broadband fluxes, observation times, and instrument frequencies. As output, it produces samples from the posterior distribution for the viewing angle, $\theta_v$, the isotropic kinetic energy released by the blastwave, $E_{\mathrm{K,iso}}$, jet core opening angle, $\theta_c$, circumburst density, $n$, the spectral slope of the electron distribution, $p$, the fraction of shock energy imparted to electrons, $\epsilon_e$, and to the magnetic field, $\epsilon_B$.

For initial prior parameters we use the best fit parameters reported in K20. The assumed prior distributions and bounds for each parameter can be viewed in detail in Table 3 of \citet{Ryan2020}. We assume a log-uniform prior distribution for $E_{\mathrm{K,iso}}$, $n$, $\epsilon_E$, $\epsilon_B$, and $\xi$ and a uniform prior distribution for $\theta_c$, $\theta_w$, and $p$. The prior distribution for the viewing angle, $\theta_v$, is constrained by the posterior probability distribution reported in \citet{Abbott2017}.

\subsubsection{Top-Hat Jet Model\label{sec:tophat}}

We begin by first calculating the simplest model which could describe the outflow geometry, the top-hat. In this scenario, the energy of the jet is independent of angle and there is an instantaneous cutoff in energy at the jet edge: 
\begin{equation}
        E(\theta)=
        \begin{cases}
            E_0 & \theta \leq \theta_{c} \\
            0 & \theta > \theta_{c}
        \end{cases}
    \end{equation}
It is unlikely that a top-hat jet could be viewed far off-axis without a significant change in the appearance of the GRB afterglow \citep{VanEerten2010, Zhang2015, Kathirgamaraju2016}. Until recently, the top-hat model was assumed for most GRB analyses. 

\begin{figure*}
\centering
\includegraphics[width=0.9\linewidth]{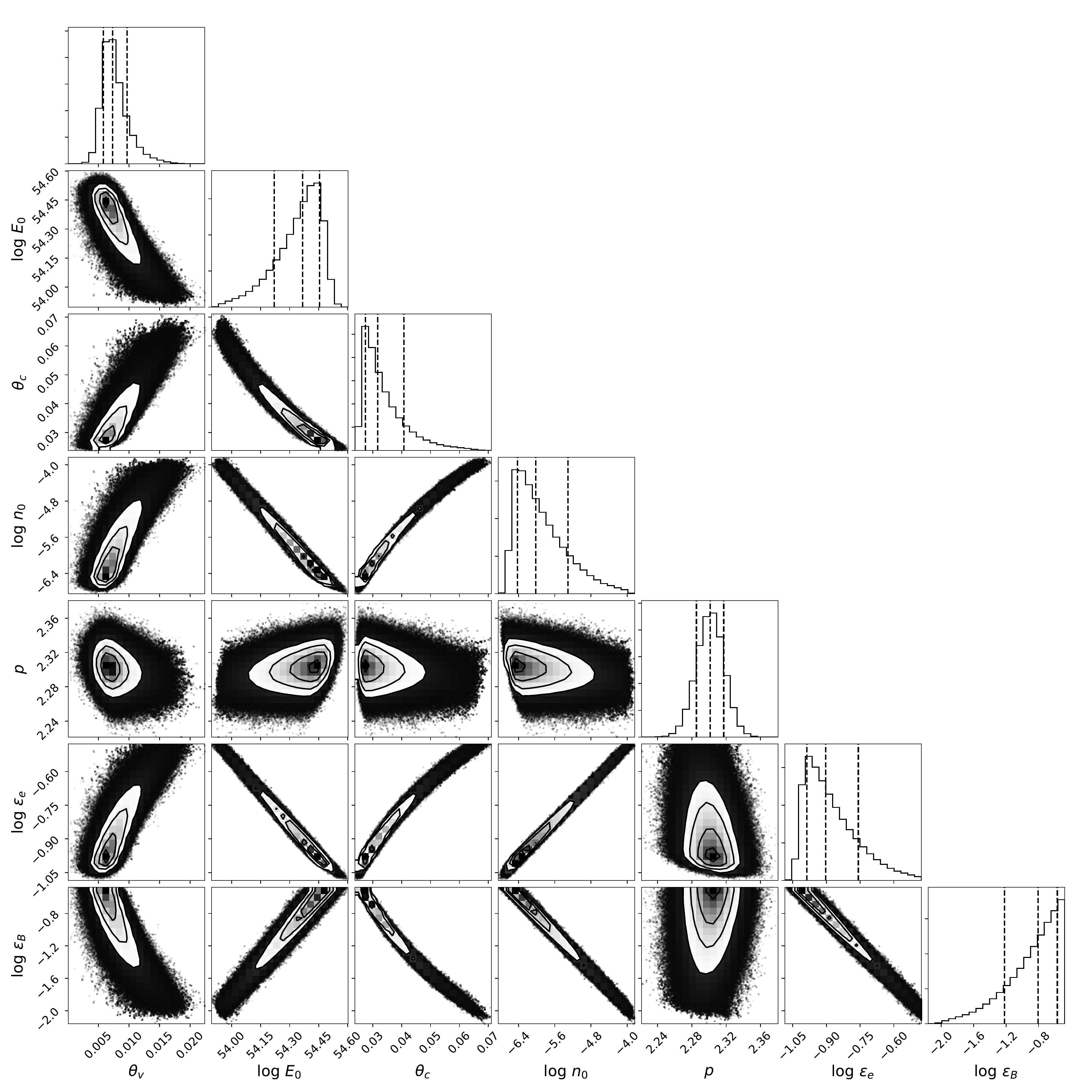}
\caption{The covariances and posterior probability distributions of the parameters for the top-hat model (\S \ref{sec:tophat}). The histograms denote the 15, 50, and 85 percentiles of the distributions.
\label{fig:corner_top}}
\end{figure*}

\begin{figure*}
\includegraphics[trim=50 0 90 0,clip,width=\linewidth]{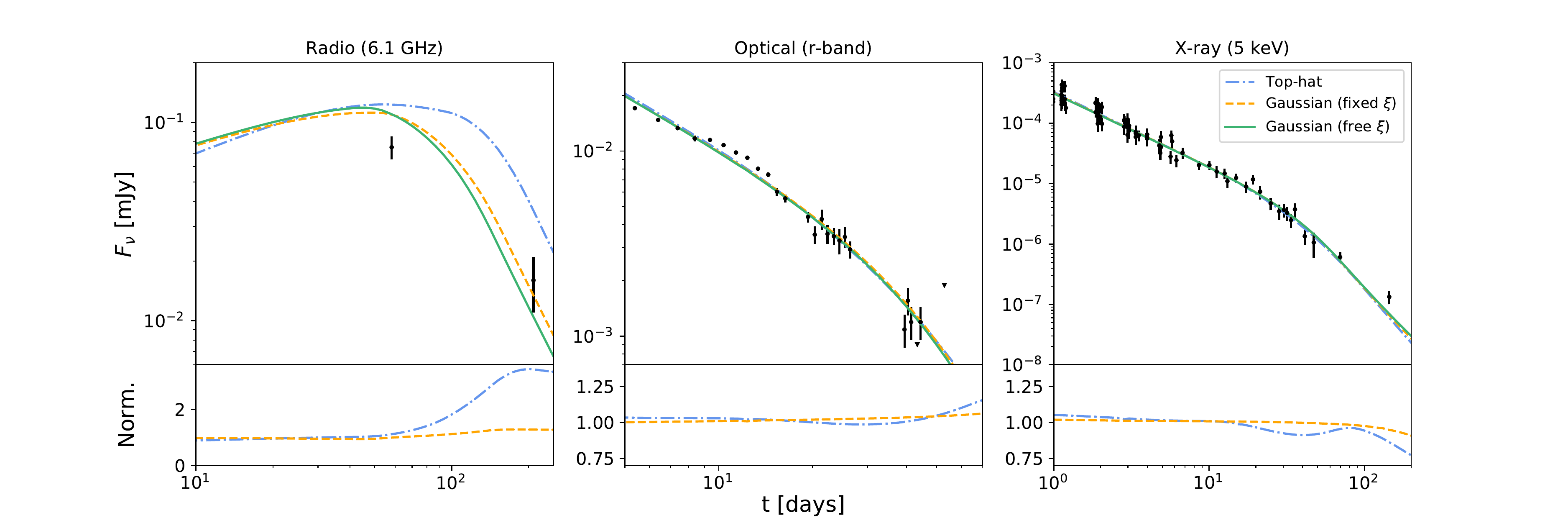}
\caption{Top panel: Comparison between each of the three models described in $\S$\ref{sec:afterglowpy}: a top-hat jet (dash-dotted line), a Gaussian jet with $\xi$ fixed to 1.0 (dashed line), and a Gaussian with $\xi$ free to vary (solid line). Bottom panel: The top-hat and fixed-$\xi$ Gaussian jet models are normalized by the free-$\xi$ Gaussian jet model. The models provide comparable fits to the observed data except at later times in the radio.
\label{fig:comp_model}}
\end{figure*}

 The covariances and posterior probability distributions of the various parameters are shown in Figure \ref{fig:corner_top}. Our best fit parameters for this model are listed in Table \ref{tab:par} and plotted in Figure \ref{fig:comp_model}. We allow the value of $p$ to vary but restrict $\epsilon_e < \nicefrac{1}{3}$ and $\epsilon_B < \nicefrac{1}{3}$ \citep{Alexander2017, Laskar2015}. This is done primarily to remove degeneracies within the model fitting; however, we found that if we do not apply the restrictions on these microphysical parameters then the fits tend towards quite unphysical values ($\epsilon_B$ approaches 1.0). Therefore, we see this as further evidence that placing these restrictions is valid in this case. The parameters derived here imply a beaming-corrected kinetic energy of $E_{K} = E_{K,\mathrm{iso}}(1-\mathrm{cos} \ \theta_{c}) = 1.2^{+0.2}_{-0.1} \times 10^{51}$ erg. 

\subsubsection{Gaussian Jet Model \label{sec:Gaussian}}

To probe the jet structure we compare the simple top-hat to a more complex Gaussian model:  
\begin{equation}
        E(\theta)=
        \begin{cases}
            E_0 \ e^{ -\frac{\theta^2}{2\theta_c^2}} & \theta \leq \theta_{w} \\
            0 & \theta > \theta_{w}
            \label{eq:gauss}
        \end{cases}
    \end{equation}
where $\theta_w$ is the truncation angle of the Gaussian wings.

\begin{figure*}
\centering
\includegraphics[width=0.9\linewidth]{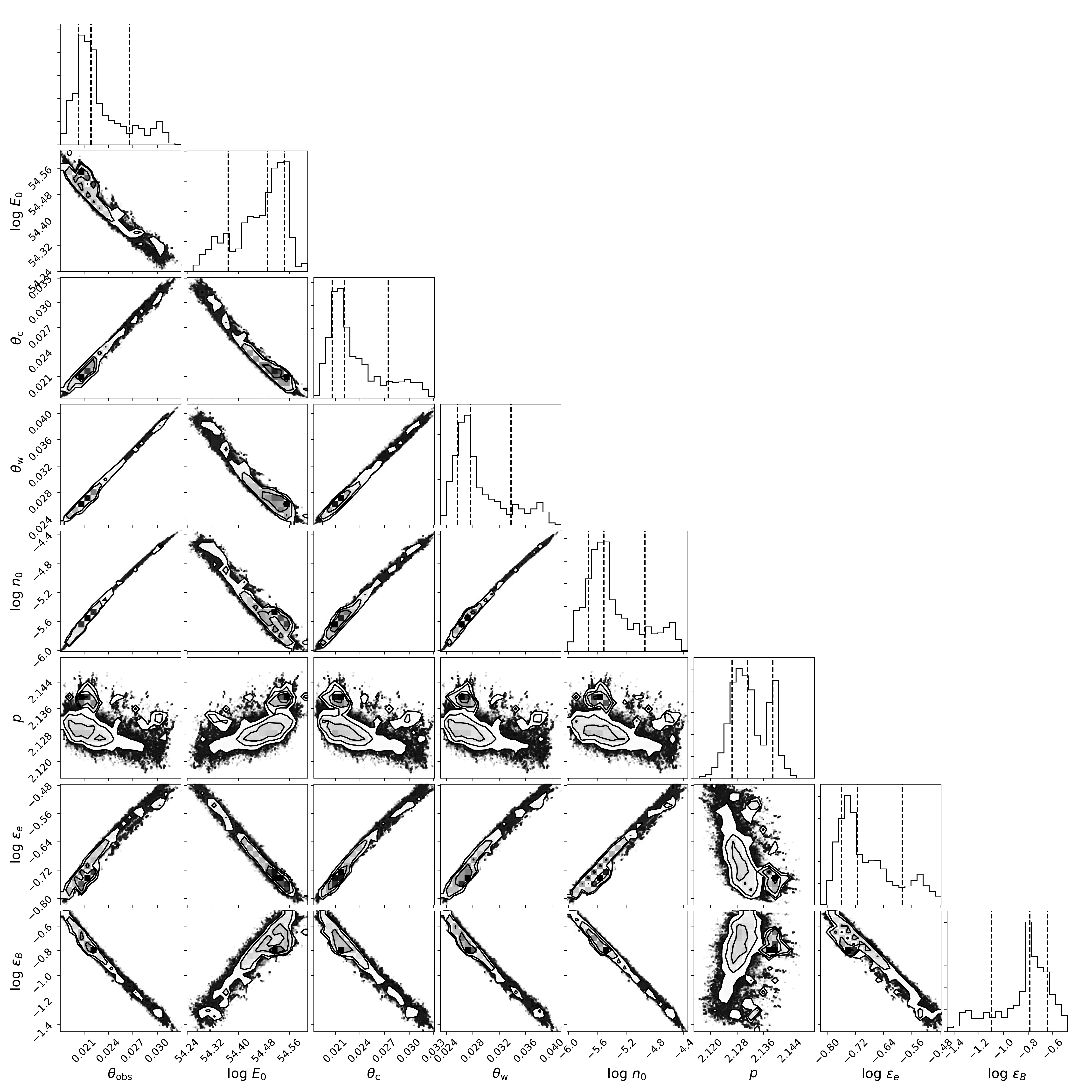}
\caption{The covariances and posterior probability distributions of the parameters for the Gaussian model with $\xi$ fixed to 1.0 (\S \ref{sec:Gaussian}). The histograms denote the 15, 50, and 85 percentiles of the distributions.
\label{fig:corner_gauss}}
\end{figure*}

Similarly to the top-hat model we again restrict $\epsilon_e < \nicefrac{1}{3}$ and $\epsilon_B < \nicefrac{1}{3}$. In Gaussian models extended emission from the jet could be viewed at angles beyond $\theta_c$ and so larger values of $\theta_v$ are possible. The best fit parameters are listed in Table \ref{tab:par} and plotted in Figure \ref{fig:comp_model}. The covariances and posterior probability distributions of the parameters are shown in Figure \ref{fig:corner_gauss}.  

To calculate the beaming-corrected energy we integrate Equation \ref{eq:gauss} over both jets:
\begin{equation}
    E = 2 \int^{\theta_w}_{0} \int^{2\pi}_{0} d\theta \ d\phi \ \text{sin}\theta \ \frac{E(\theta)}{4\pi},
\end{equation}

which approximates to:
\begin{equation}
    E \sim E_{0} \ \theta_{c}^2 \left(1 -  e^{\frac{-\theta_w^2}{2\theta_c^2}} \right).
    \label{eq:gauss_etot}
\end{equation}
This gives a value for the beaming-corrected kinetic energy of $E_K = 8.4^{+1.2}_{-0.7} \times 10^{50}$ erg, which is slightly less than that found for the top-hat model. 

\subsubsection{\texorpdfstring{Gaussian Jet Model with Free $\xi$ \label{sec:xi}}{Gaussian Jet Model with Free Xi}}

In the synchrotron afterglow model the emission is driven by a power-law distribution of electrons in the surrounding medium. The participation fraction, $\xi$, describes the percentage of total electrons which are accelerated by the passing shock wave and contribute to this power-law distribution. 100\% participation ($\xi=1$) is typically assumed in the literature but simulations have shown $\xi$ can be as low as $10^{-2}$ \citep{Sironi2011, Sironi2013} and that lower values tend to be more realistic \citep{Warren2018}.

To test this we expand on our Gaussian model and now allow $\xi$ to vary as a free parameter (Table \ref{tab:par}). Notably, the beaming-corrected kinetic energy in this scenario is $E_K = 1.1^{+6.5}_{-0.9} \times 10^{53}$ erg, two orders of magnitude higher than in the previous cases. Such a high energy density may lead to concerns over the ability of the afterglow emission to avoid becoming suppressed by processes such as pair-production opacity. However, at the later times described here the afterglow has had sufficient time to expand and become diffuse. For an X-ray photon the opacity due to pair production is quite low ($\sim 10^{-8}$). This is because the number of high energy (GeV) photons with which the X-ray photon could pair-produce is small and so it is free to travel unhindered through the shock wave. During the prompt emission and possibly for GeV afterglows this could be a bigger concern, but for the later, more diffuse afterglow we do not consider it an issue.

\begin{figure*}
\centering
\includegraphics[width=0.9\linewidth]{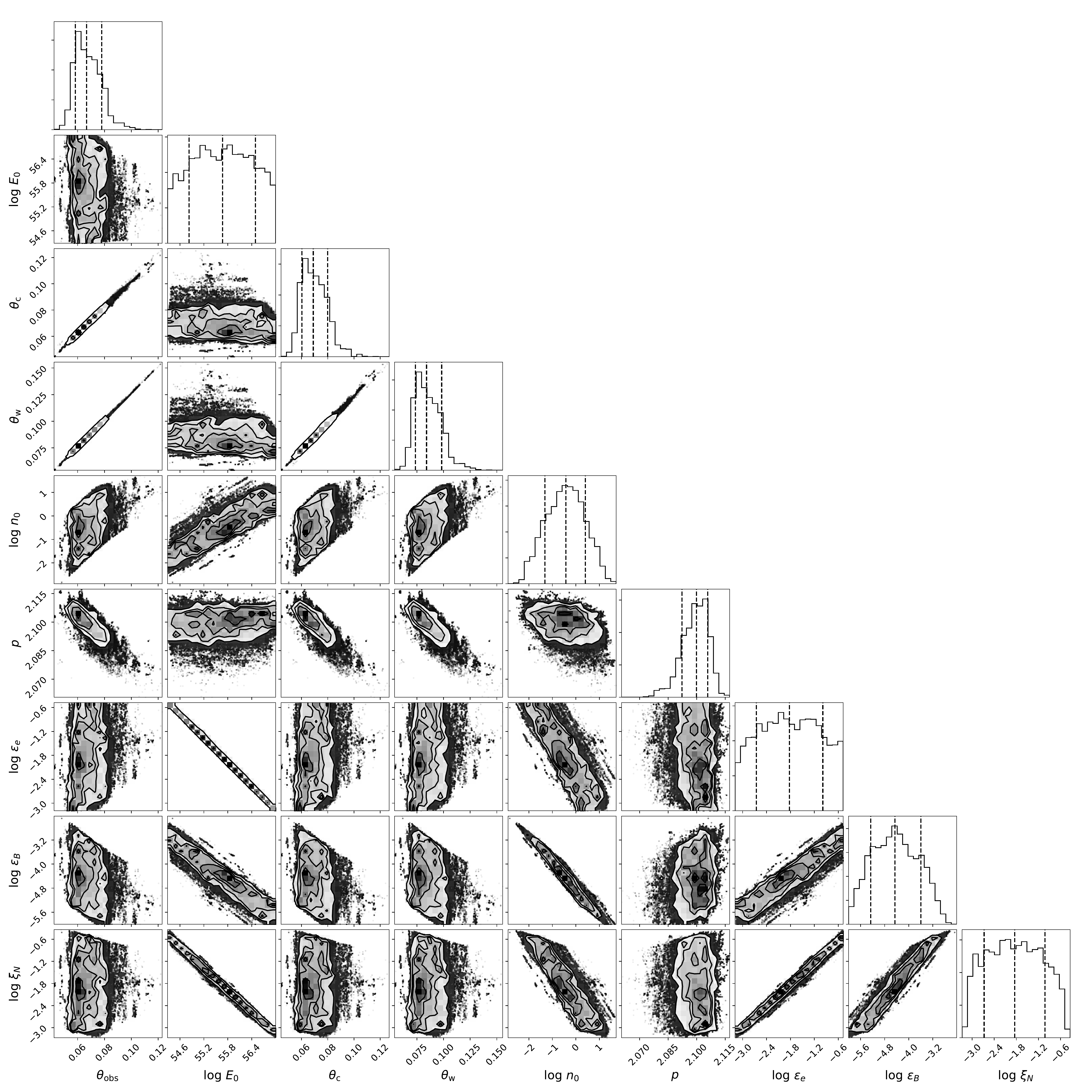}
\caption{The covariances and posterior probability distributions of the parameters for the Gaussian model with $\xi$ left to vary freely (\S \ref{sec:xi}). The histograms denote the 15, 50, and 85 percentiles of the distributions.
\label{fig:corner_gauss_xi}}
\end{figure*}

\begin{figure*}
\includegraphics[trim=40 10 60 50,clip,width=1.0\linewidth]{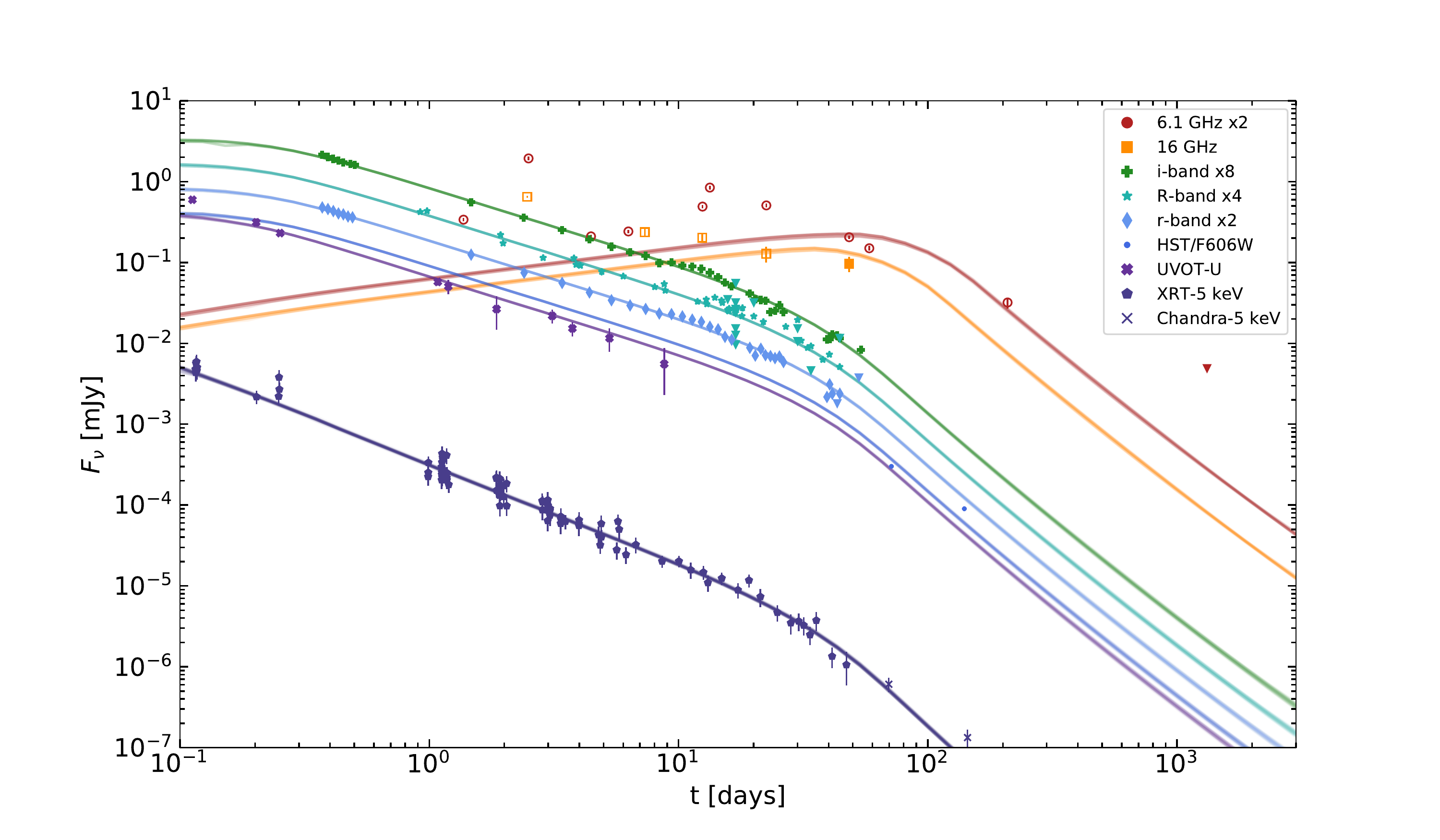}
\caption{The multiwavelength light curve of GRB\,160625B overplotted with our best fit model of the afterglow forward shock from $\S$\ref{sec:xi}: a Gaussian jet with $\xi$ free to vary. Times are referenced from the GBM trigger (Jun 25 2016 22:40:16.28 UT). Open points represent radio data which were available but not included in the analysis (See $\S$\ref{sec:tenets} for more details). Transparent lines represent model uncertainties and are calculated via samples taken from the MCMC posterior distribution.
\label{fig:lc}}
\end{figure*}

We discuss this case in more detail in $\S$\ref{sec:xidisc}. The covariances and posterior probability distributions of the parameters are shown in Figure \ref{fig:corner_gauss_xi}. We directly compare the previous top-hat and Gaussian jet models to this case in Figure \ref{fig:comp_model} and plot the best fit over the data in Figure \ref{fig:lc}. 

\begin{figure*}
\includegraphics[trim=50 0 90 0,clip,width=\linewidth]{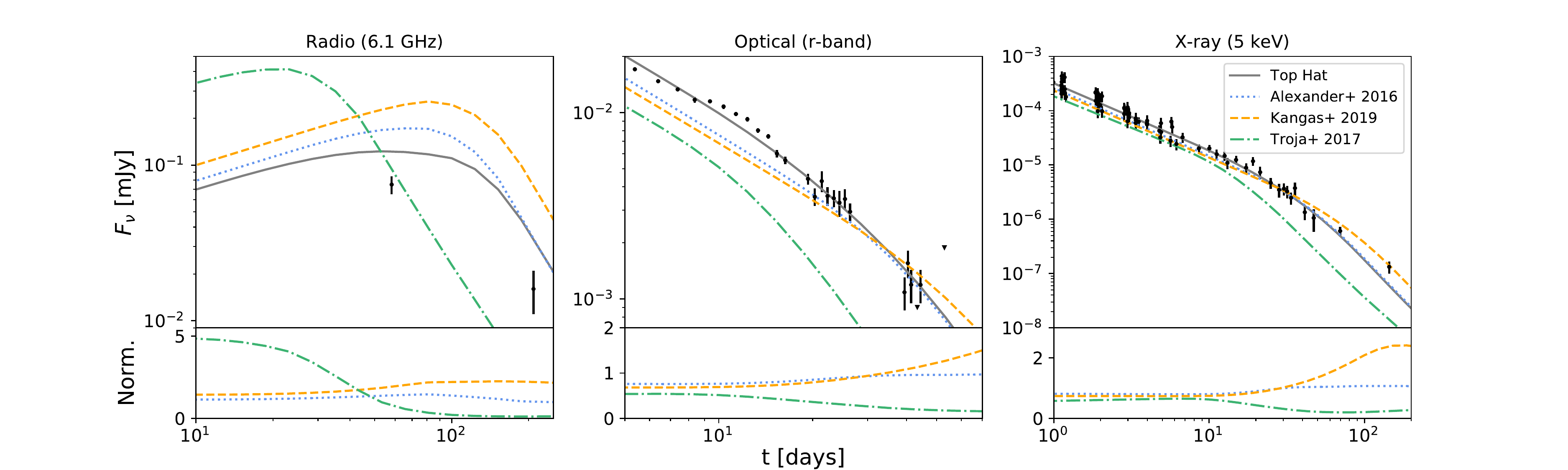}
\caption{Top panel: Comparison between our top-hat jet model (solid line) to previous works from A17 (dotted line), K20 (dashed line), and T17 (dash-dotted line) (Table \ref{tab:par}). Bottom panel: The A17, K20, and T17 models are normalized by our top hat model ($\S$\ref{sec:tophat}). At X-ray wavelengths the models are almost indistinguishable at early times. The discrepancies between models are more apparent at radio and optical frequencies, where fewer observations are available. 
\label{fig:comp_struc}}
\end{figure*}

\section{Discussion} \label{sec:discussion}

\subsection{Comparison to Past Works}

Previous works have completed similar analyses on GRB\,160625B (T17, A17, K20). In each instance the jet is modeled with a conical top-hat structure and there is general agreement on the burst parameters (Table \ref{tab:par}). We begin by simulating the light curves with \texttt{afterglowpy} using the model parameters found in A17, T17, and K20. We then compare the results to our own top hat solution in Figure \ref{fig:comp_struc}.

K20 utilize the \texttt{BoxFit} package \citep{vanEerten2012} to model the afterglow while A17 base their results on the synchrotron model described in \cite{Laskar2014, Laskar2015}. In Figure \ref{fig:comp_struc} the differences between the models at radio, optical, and X-ray wavelengths can be primarily explained by systematic offsets between \texttt{afterglowpy} and the other models used, differences in the datasets used, and the use of additional late-time data that was not available yet for A17 and T17. The largest discrepancies are seen at radio energies. This is partially due to the fewer number of observations available and the potential contamination of the forward shock by other radio effects. Recently \citet{Kangas2019} have noted inconsistencies in observed jet break times between radio and higher frequencies; thus suggesting that radio afterglow light curves may simply not be well represented by standard afterglow theory. \citet{Jacovich2020} attribute this discrepancy to the lack of proper implementation of Klein-Nishina and effects in most afterglow modeling codes.





\subsection{Model Comparison with WAIC}

To quantify the differences between our own top-hat and Gaussian models described in $\S$\ref{sec:afterglowpy} we utilize the Widely Applicable Information Criterion (WAIC; \citealt{Watanabe2010, Troja2020}). The WAIC score provides an estimate of the expected log predictive density (elpd), i.e., how likely the model is to provide a good fit for future data \citep{Gelman2013}. The elpd in general is quite hard to derive without prior knowledge of the true model but the WAIC score can be calculated directly from the MCMC statistical samples. In general, it is the difference between WAIC scores, rather than the raw WAIC score itself, which is most relevant. A model is considered strongly preferred, i.e., has greater predictive power, over another if the difference between their two WAIC scores is a factor of a few larger than the error on that difference. The uncertainty on the raw WAIC and WAIC difference scores is an estimate of the standard error and can be an underestimate but is usually accurate within a factor of 2 \citep{Bengio2004}. Therefore we list a significance range for the confidence level in Table \ref{tab:modelcomp}.

As discussed in $\S$\ref{sec:afterglowpy} we first began by directly comparing a top-hat and Gaussian style jet and then exploring the effects that varying the participation fraction, $\xi$, had on the GRB afterglow. Table \ref{tab:modelcomp} shows the model comparison between each of these three cases. Both Gaussian models show a greater predictive power compared to the simpler top-hat model but there is not a significant difference between the two Gaussian models themselves.

\subsection{\texorpdfstring{The Participation Fraction, $\xi$ \label{sec:xidisc}}{The Participation Fraction}}

In $\S$\ref{sec:xi} we fit the afterglow of GRB\,160625B with a Gaussian jet model but allowed the participation fraction, $\xi$, to vary. In agreement with the findings of \citet{Warren2018} our model prefers a lower value of \mbox{$\xi$$\sim$0.01}, rather than total participation. Clearly, decreasing the participation fraction has dramatic effects on the other physical parameters. If all other parameters are kept fixed then the density of the circumburst environment must increase to provide the necessary number of electrons to produce the observed flux. This causes $\theta_c$ to increase as the relativistic jet interacts with more material. Increasing the density also results in a faster-evolving shock wave, so $E_{K,\mathrm{iso}}$ must increase to maintain the light curve shape. Most notably, $\epsilon_B$ decreases by four orders of magnitude due to the lack of accelerated electrons.  

Figure \ref{fig:xi} illustrates how varying $\xi$ can have dramatic effects on the predicted afterglow light curve. We begin with the best fit parameter values of the top-hat jet and Gaussian (fixed-$\xi$) jet models from $\S$\ref{sec:afterglowpy} but fix $\xi$ to three different values $-$ 1.0, 0.1, and 0.01 $-$ and plot the results. We see the greatest differences at radio energies and at early times in the optical band. Those electrons which pass through the shock wave without being accelerated may increase the opacity to synchrotron self-absorption and also introduce an additional source of emission at very early times at optical wavelengths ($\sim$few seconds post-burst) that then remains detectable at radio/millimeter wavelengths for several days or more \citep{Ressler2017}. Therefore, constraining the value of $\xi$ is critical for understanding the implications of the total energy budget of the burst ($\S$\ref{sec:energy}) and the density of the local circumburst environment ($\S$\ref{sec:n}).  
\subsection{Sharp Edge Effects on p \label{sec:sharpedge}}

In both Gaussian jet fits the spectral slope of the electron distribution, $p$, is significantly lower than $\sim$2.3 which is the value found both analytically ($\S$\ref{sec:tenets}) and in the top-hat model case ($\S$\ref{sec:tophat}). When $\xi$ is allowed to vary $p$ decreases from $2.13\pm0.01$ to $2.10\pm0.01$. This is explained by the favoured relationship between $\theta_c$ and $\theta_w$. In both Gaussian models we find $\nicefrac{\theta_c}{\theta_w} \sim 0.8$ meaning the emission does not extend greatly off-axis beyond the primary portion of the jet. Therefore, in slightly off-axis viewing angles the effect of a sharp edge may have a significant impact on the resulting light curve. Emission from one side of the jet reaches the viewer before the other side, resulting in slightly less observed flux than expected for a perfectly on-axis viewing angle. This manifests as a steepening of the light curve which can then allow $p$ to instead probe lower values in the parameter distribution. 

To investigate this we repeated the fit from \ref{sec:xi} but now place a constraint where $\theta_w > 3\theta_c$ so as to force a `softer' edge to the jet. In this scenario we find that $p$ prefers a higher value of $2.23^{+0.02}_{-0.03}$, more consistent with the top-hat jet results, and $\theta_w = 0.59^{+0.32}_{-0.16}$ (Table \ref{tab:varying}). All other parameters remain consistent with those of the free-$\xi$ case in Table \ref{tab:par}. A softer-edged jet may represent a more physically realistic scenario compared to a sudden drop-off in emission at the jet edge. However, a WAIC analysis between this case (WAIC = $799.0 \pm 126.8$) and the freely varying $\xi$ case (WAIC = $1782.7 \pm 79.1$) shows better predictability for the sharp-edged case. This is, in fact, consistent with the results of \citet{BeniaminiEhud2019} where the authors claim that emission from structured jets cannot be observed far from the core of the jet.


\subsection{Viewing Angle Effects \label{sec:viewingangle}}

Precise predictions for GRB viewing angles have only become possible within the last few years thanks to the advent of various powerful high-resolution hydrodynamic simulations used both directly and indirectly by codes such as \texttt{BoxFit} \citep{vanEerten2012}, \texttt{ScaleFit} \citep{Ryan2013}, \texttt{JET} \citep{Duffell2013a}, and \texttt{afterglowpy} \citep{Ryan2020}. \citet{Ryan2015} show that, in fact, most GRBs are probably observed off-axis and the joint discovery of GW\,170817/GRB\,170817A highlighted just how significant viewing angle effects could be for a single event \citep{Lazzati2018, Troja2018, Xie2018, Alexander2018, Wu2018, Fong2019, Troja2019MNRAS, Lamb2020}.

Underestimating the viewing angle of GRBs may introduce biases in the afterglow model fitting, e.g., by overestimating the beaming width of the jet \citep{Vaneerten2015}. To illustrate this we repeat the Gaussian model with the same initial conditions as described in $\S$\ref{sec:Gaussian} but now we fix $\theta_v$ to an almost on-axis angle of $10 ^{-4}$ radians (Table \ref{tab:varying}). As a reference we note that \swift\ GRBs are thought to be observed more off-axis than this. Typical values of $\theta_v$ range from 0.055 to 0.42 radians \citep{Ryan2015}.

In this case we find that by assuming an on-axis viewing angle we are overestimating $\theta_c$ by a factor of $\sim$4.1 (compared to the fixed-$\xi$ Gaussian jet model). Utilizing Equation \ref{eq:gauss_etot} the beaming-corrected kinetic energy in this case is $E_K = 2.0^{+0.4}_{-0.1} \times 10^{51}$ erg. Although we have overestimated the beaming angle the slightly lower estimate for $E_{\mathrm{K,iso}}$ means the beaming-corrected kinetic energy remains consistent with the Gaussian case. Interestingly, the fact that $\nicefrac{\theta_w}{\theta_c} < 1$ indicates a strong preference for a sharp jet edge as in the case of a top-hat jet. However, a WAIC comparison between this case (WAIC = 1145.7$\pm$197.1) and the Gaussian jet with fixed-$\xi$ (WAIC = 1744.3$\pm$78.5) still shows better predictability and preference for a Gaussian-shaped jet model.


\subsection{Local Circumburst Environment \label{sec:n}}

Long GRBs are believed to result from the end state of massive stars \citep{Woosley2006}. Due to their short lifespan (tens of Myr), these massive stars live and die in the same dense molecular clouds which birthed them. It would appear to be a reasonable assumption to expect the local circumburst densities of GRBs to reflect that of regions of high star-formation ($> 1 \ \mathrm{cm}^{-3}$). Observationally this is not usually the case, at least when assuming a top-hat jet model and $\xi$=1 \citep{Laskar2015}. 

The derived value of the local circumburst density, $n$, for GRB\,160625B is exceptionally low for most models in Table \ref{tab:par}. This parameter can be difficult to constrain as it is highly degenerate with other physical parameters in the system. For most GRBs the value of $n$ tends to be between $\sim10^{-3}$ and $\sim10^2$ cm$^{-3}$ when assuming $\xi$=1 \citep{Laskar2015}. Density measurements between \swift\ and \fermi\ GRB populations tend to occupy discrete regions of parameter space, thus leading previous studies (e.g., \citealt{Racusin2011}) to suggest that these two populations may originate in different host environments, although there was not a large enough sample at the time to definitively confirm this.

The preference for ISM-like, low-density local environments may suggest LAT-detected long GRBs originate from lower-metallicity massive stars. This is motivated by the fact that these types of progenitor stars tend to have lower mass-loss rates \citep{Kudritzki1987, Vink2001, Woosley2002}. The ability for the relativistic jet to travel unhindered may also prevent the suppression of several radio components and can allow the reverse shock to propagate freely. GRB\,160625B is one of only a few long-duration GRBs with a confirmed reverse shock (e.g., \citealt{Laskar2013, Perley2014, Laskar2016, Laskar2018, Laskar2019}). 

In the standard afterglow model the circumburst density is intricately connected to other observed physical parameters. As noted in the case of the Gaussian jet with free-$\xi$ (\S\ref{sec:xi}) our estimate of the local circumburst density is highly dependent upon the participation fraction. Decreasing $\xi$ by a factor of 100 can increase $n$ by upwards of five orders of magnitude. Therefore, further work on the impact of the participation fraction, $\xi$ is required before a definitive association can be made between highly energetic long GRBs and massive, metal-poor progenitor stars.

\begin{figure*}
\includegraphics[trim=50 0 90 0,clip,width=\linewidth]{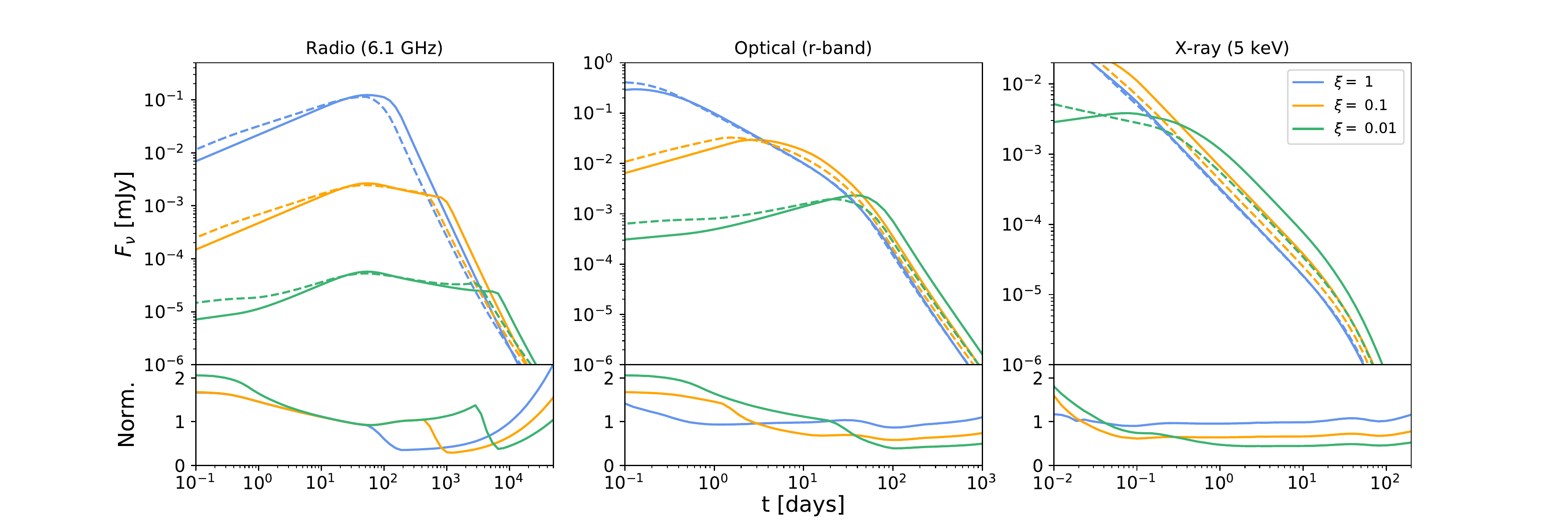}
\caption{Top panel: We present the best fit parameter values of the top-hat (solid line) and Gaussian (dashed line) jet models from $\S$\ref{sec:afterglowpy} but at three fixed values of $\xi$. The participation fraction, $\xi$, plays a pivotal role in the GRB afterglow, especially at radio energies. Lowering the value of $\xi$ can have dramatic effects on the light curve. To account for the lower number of electrons contributing to the emission the total relativistic energy released must be increased and/or the density of the local circumburst environment must increase as well. Bottom panel: The Gaussian fluxes are normalized by the top-hat model for each value of $\xi$.
\label{fig:xi}}
\end{figure*}



\begin{table*}
  \centering
 \caption{Best Fit Model Parameters. }
  \begin{tabular}{cccccccc}
    \toprule
     & & This Work & This Work & This Work & A17 & K20 & T17 \\
Model & & Top-Hat & Gaussian (fixed $\xi$) & Gaussian (free $\xi$) & Top-Hat & Top-Hat & Top-Hat \\ \hline

$\theta_v$ & [rad] & 0.0073$^{+0.0024}_{-0.0015}$ & 0.022$^{+0.006}_{-0.001}$ & 0.066$^{+0.010}_{-0.009}$ & - & 0.012 & - \\

$E_{\text{K,iso}}$ & [erg] & $2.3^{+0.5}_{-0.7} \times 10^{54}$ & $3.1^{+0.4}_{-0.8} \times 10^{54}$ & $4.2_{-3.5}^{+24} \times 10^{55}$ & $1.1^{+1.0}_{-0.5} \times 10^{54}$ & $1.8 \times 10^{54} $ & $2.0^{+1.0}_{-1.4} \times 10^{54}$\\

$\theta_c$ & [rad] & 0.032$^{+0.009}_{-0.004}$ & 0.022$^{+0.006}_{-0.001}$ & 0.068$^{+0.010}_{-0.010}$ & 0.063 $\pm$ 0.003 & 0.059 & 0.042$^{+0.028}_{-0.012}$\\

$\theta_w$ & [rad] & - & 0.028$^{+0.007}_{-0.002}$ & 0.083$^{+0.014}_{-0.011}$ & - & - & -\\

$n$ & [cm$^{-3}$] & $9.6_{-5.9}^{+39.0} \times 10^{-7}$ & $3.1_{-1.1}^{+11} \times 10^{-6}$ & 0.352$^{+1.71}_{-0.307}$ & $(5\pm3)\times 10^{-5}$ & $1.1 \times 10^{-5}$ & $1^{+49}_{-0.9} \times 10^{-4}$\\

$p$ & & 2.30$\pm$0.02 & 2.13$^{+0.01}_{-0.01}$ & 2.10$^{+0.01}_{-0.01}$ & 2.31 $\pm$ 0.01 & 2.30 & 2.2\\

$\epsilon_e$ & & 0.12$_{-0.02}^{+0.05}$ & 0.19$_{-0.02}^{+0.07}$ & 0.017$^{+0.085}_{-0.014}$ & $0.23^{+0.07}_{-0.08}$ & 0.13 & $0.10^{+0.22}_{-0.09}$ \\

$\epsilon_B$ & & 0.16$_{-0.10}^{+0.11}$ & 0.17$_{-0.09}^{+0.05}$ & $3.9^{+21.9}_{-3.3} \times 10^{-5}$ & $0.013_{-0.01}^{+0.11}$  & 0.030 & $0.010^{+0.090}_{-0.009}$ \\

$\xi$ & & 1.0 & 1.0 & 0.016$^{+0.080}_{-0.013}$ & 1.0 & 1.0 & 1.0 \\

$\eta \ ^{\mathrm{a}}$ & & $0.56^{+0.08}_{-0.05}$ & $0.49^{+0.08}_{-0.03}$ & $0.067^{+0.253}_{-0.057}$ & $0.73^{+0.10}_{-0.14}$ & 0.62 & $0.75^{+0.16}_{-0.09}$ \\

$E_{\text{rel}}\ ^{\mathrm{b}}$& [erg] & $2.7^{+1.3}_{-0.5} \times 10^{51}$ & $1.6^{+0.6}_{-0.1} \times 10^{51}$ &
$1.2^{+6.5}_{-0.9} \times 10^{53}$ &
$2.3^{+1.8}_{-1.2} \times 10^{51}$ & 
$8.3 \times 10^{51}$ & $\sim6 \times 10^{51}$ \\

$\nicefrac{\chi^2}{\text{dof}}$ & & 1.24 & 0.99 & 0.86 & 1.26 & 8.6 & -\\
\hline
  \end{tabular}
   \begin{tablenotes}
      \small
      \centering
      \item Uncertainties are given at 1$\sigma$
      confidence levels. The reduced-$\chi^2$ is the minimum over all completed runs.
      \item $^{\mathrm{a}}$ $\eta = \frac{E_{\gamma,\mathrm{iso}}}{E_{\gamma,\mathrm{iso}} + E_{\text{K,iso}}}$, assuming $E_{\gamma,\text{iso}} \sim 3 \times 10^{54}$ erg \citep{Zhang2018}
      \item $^{\mathrm{b}}$ $E_{\text{rel}} = E_{\gamma} + E_{\text{K}}$
    \end{tablenotes}
  \label{tab:par}
\end{table*}

\subsection{Energetics and Central Engine \label{sec:energy}}

One of the main goals of this and future work will be calculating the beaming-corrected energy released by the GRB and especially how that relates to the physics of the central engine. Currently, two popular progenitor models are favored to explain the central engines of long GRBs: magnetars and black hole systems. 

The spindown of a newborn millisecond pulsar could potentially power a long GRB via a Poynting-flux dominated relativistic outflow \citep{Zhang2001,Thompson2004}. These magnetars are limited by their finite reserve of rotational energy. For a neutron star with a mass limit of $\sim2 \ M_{\odot}$, a radius of 10 km, and a spin period of about 1 ms places a cap on the available energy at $\sim10^{52}$ erg \citep{Metzger2007, Metzger2011}. This energy cap can be increased in the case of `supramassive' neutron stars that have been stabilized by centrifugal forces and therefore could accommodate rotational energies of up to $10^{53}$ erg \citep{Metzger2015}. However, \cite{Beniamini2017} and \cite{Metzger2018} have recently shown that the true reservoir of available energy is unlikely to reach beyond ${\sim few \times 10^{52}}$ erg and so the magnetar model is in strong tension with at least the most energetic long GRBs.


In the second scenario, a black hole is formed in the immediate aftermath of the death of a massive star \citep{Woosley1993,Macfadyen1999,Woosley2012}. Stellar material accreting back onto the newly formed black hole may power relativistic jets. The jets burrow through the stellar envelope and provide an outlet for relativistic material to escape \citep{Morsony2007}. The typical 10-100 s durations of long GRBs correspond to the free-fall time of the star's helium core. Unlike magnetars, these `collapsars' have much less stringent caps on the potential energy which could be extracted. Theoretical predictions can easily produce energy releases greater than $10^{53}$ erg. 

Previous estimates of the total relativistic energy ($E_{\text{rel}} = E_{\gamma} + E_{\text{K}}$) produced by GRB\,160625B range from (2.3-8.3)$\times 10^{51}$ erg when assuming $\xi$=1 (Table \ref{tab:par}). This is approaching but still within the upper limit of the magnetar model. We found in \S\ref{sec:Gaussian} that modeling the jet as a Gaussian rather than a top-hat leads to an estimate of the total relativistic energy that is below this range. Clearly, based on energetic arguments alone,  we cannot rule out either the magnetar or collapsar model as progenitors. However, as noted in \S\ref{sec:xi}, assuming the participation fraction of electrons is unity is unlikely to be realistic. In agreement with K20, we find that more reasonable lower values of $\xi$ produce an estimate of the total relativistic energy which is two orders of magnitude higher (Figure \ref{fig:xi}). This value can still be reasonably accommodated by the collapsar model but is in strong tension with the energy cap expected from a magnetar. Regardless, it serves to show that without a better understanding of the participation fraction it will be difficult to draw robust conclusions regarding the progenitor.

\begin{table*}
  \centering
 \caption{Best Fit Gaussian Model Parameters for $\S$\ref{sec:sharpedge} and $\S$\ref{sec:viewingangle}}
  \begin{tabular}{cccc}
    \toprule
 & & Soft Jet Edge & Fixed $\theta_v$  \\
$\theta_v$ & [rad] & $0.134^{+0.042}_{-0.039}$ & $10^{-4}$ \\

$E_{\text{K,iso}}$ & [erg] & $1.1^{+1.6}_{-0.7} \times 10^{55}$ & $3.9^{+0.3}_{-0.6} \times 10^{54}$  \\

$\theta_c$ & [rad] & $0.093^{+0.030}_{-0.026}$ & $0.09\pm0.02$ \\

$\theta_w$ & [rad] & $0.59^{+0.32}_{-0.16}$ & 0.024$^{+0.003}_{-0.001}$ \\

$n$ & [cm$^{-3}$] & $4.1^{+25.2}_{-3.7}$ & $3.4_{-1.0}^{+4.6} \times 10^{-7}$ \\

$p$ & & $2.23^{+0.02}_{-0.03}$ & $2.301^{+0.003}_{-0.005}$ \\

$\epsilon_e$ & & $0.076^{+0.113}_{-0.045}$ & 0.079$_{-0.006}^{+0.015}$  \\

$\epsilon_B$ & & $7.0^{+18}_{-4.0}\times10^{-6}$ & 0.25$_{-0.10}^{+0.06}$ \\

$\xi$ & & $0.061^{+0.054}_{-0.029}$ & 1.0 \\

$\eta \ ^{\mathrm{a}}$ & & $0.21^{+0.20}_{-0.11}$ & $0.44^{+0.05}_{-0.02}$  \\

$E_{\text{rel}} \ ^{\mathrm{b}}$& [erg] & $1.3^{+0.9}_{-0.6} \times 10^{53}$ & $2.0^{+0.4}_{-0.1} \times 10^{51}$  \\

$\nicefrac{\chi^2}{\text{dof}}$ & & 6.28 & 7.24 \\
\hline
  \end{tabular}
   \begin{tablenotes}
      \small
      \centering
      \item Uncertainties are given at 1$\sigma$
      confidence levels. The reduced-$\chi^2$ is the minimum over all completed runs.
      \item $^{\mathrm{a}}$ $\eta = \frac{E_{\gamma,\mathrm{iso}}}{E_{\gamma,\mathrm{iso}} + E_{\text{K,iso}}}$, assuming $E_{\gamma,\text{iso}} \sim 3 \times 10^{54}$ erg \citep{Zhang2018}
      \item $^{\mathrm{b}}$ $E_{\text{rel}} = E_{\gamma} + E_{\text{K}}$
    \end{tablenotes}
  \label{tab:varying}
\end{table*}

\section{Conclusions\label{sec:conclusions}}

With this work we have performed a case study of GRB\,160625B to show the benefits of detailed multiwavelength afterglow modeling as it pertains to understanding GRB energetics and environments. We modeled observations from radio to X-ray wavelengths spanning 0.1 to 1319 days post trigger. Using the standard afterglow framework we derived values for several physical parameters pertaining to the burst and performed a comparison between top-hat and Gaussian jet structure models. Our main conclusions can be summarized as follows:

\begin{itemize}
    \item We fit GRB\,160625B with a top-hat jet via the \texttt{afterglowpy} modeling package. We find general agreement in the afterglow parameters with previous top-hat jet models.
    \item Next, we fit the afterglow with a Gaussian-shaped jet. Although the derived density, kinetic energy, and other microphysical parameters remain consistent with the top-hat case we find that this jet shape is strongly preferred. 
    \item Finally, we considered how allowing more freedom for the participation fraction, $\xi$, affects the afterglow parameters. This change had the most dramatic effect and resulted in a density which is 5 orders of magnitude higher, a value of $\epsilon_B$ which is 4 orders of magnitude lower, and a total relativistic energy which is 2 orders of magnitude higher. This has important implications for constraining the GRB local circumburst environment, central engine, and burst energetics.
\end{itemize}


\begin{figure*}
\includegraphics[trim=50 0 90 0,clip,width=\linewidth]{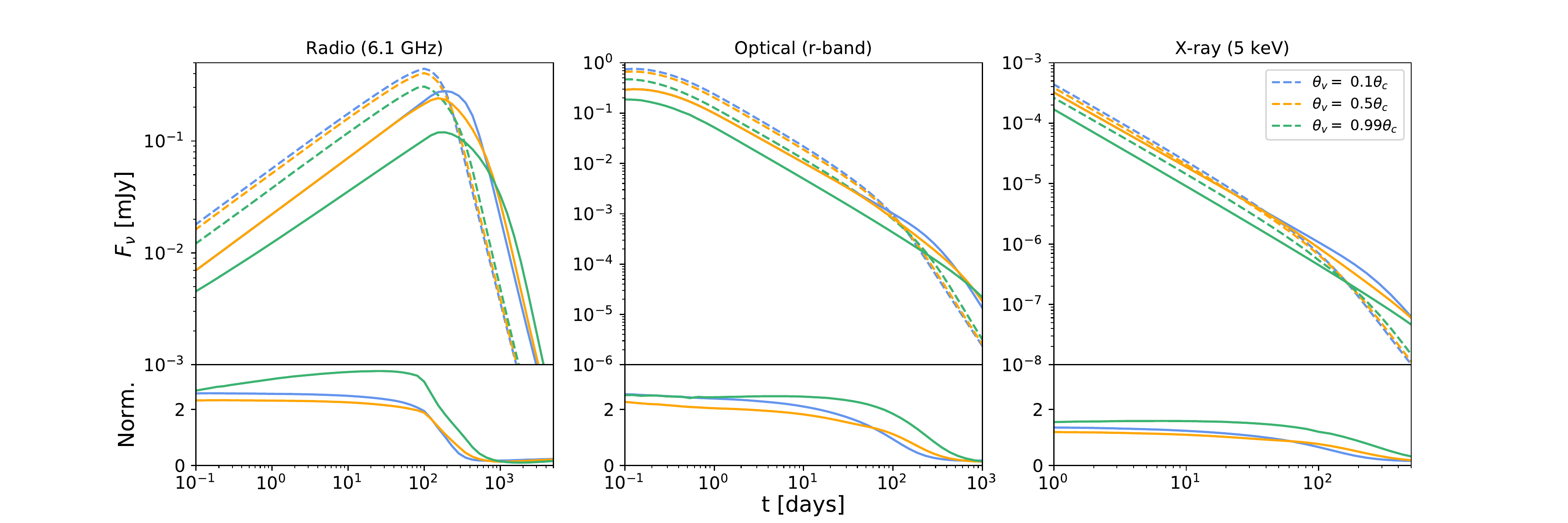}
\caption{Top panel: Here we calculate the afterglow light curves for both a top-hat (solid line) and Gaussian (dashed line) jet structure at various points within the beam. We fix $\theta_c$ in the top-hat model so that $E_{\mathrm{tot}}$ remains the same in both models. All other parameters besides $\theta_v$ and $\theta_c$ are the same as listed in Table \ref{tab:par}. Viewing a GRB off-axis has the greatest affect at early times as this is when edge effects are most noticeable. The differences between Gaussian and top-hat models are most apparent at later times.  Bottom panel: The Gaussian fluxes are normalized by the top-hat models for each viewing angle. 
\label{fig:thv}}
\end{figure*}

Given that our models include several highly degenerate parameters it can be challenging to distinguish between their various subtleties. In Figure \ref{fig:thv} we show how varying the viewing angle, $\theta_v$, impacts the burst afterglow for both top-hat and Gaussian jet structures. There exists an offset in flux density at late times between the two models that is independent of waveband or viewing angle. Therefore the detection of GRBs viewed substantially off-axis (`orphan afterglows') may not be strictly necessary and future multi-wavelength observations at very late times could help further reinforce the preference for a Gaussian-shaped jet over a top-hat jet.


In future work we plan to continue our analysis on the afterglow modeling of bright LAT-detected long GRBs. These events typically have abundant data and also tend to be observed in low-density environments, thus providing an excellent sample for comparing jet structure models. We will use the results to expand on the arguments discussed here and create a comprehensive sample of well-documented LAT long GRBs and their general properties. 

\section*{}

Analysis was performed on the YORP cluster administered by the Center for Theory and Computation, part of the Department of Astronomy at the University of Maryland. GR acknowledges the support from the University of Maryland through the Joint Space Science Institute Prize Postdoctoral Fellowship. A.C. acknowledges support from NSF CAREER award \#1455090. A.H. acknowledges support by the I-Core Program of the Planning and Budgeting Committee and the Israel Science Foundation. The National Radio Astronomy Observatory is a facility of the National Science Foundation operated under cooperative agreement by Associated Universities, Inc. We acknowledge the use of public data from the Swift data archive.

\facilities{ SWIFT (XRT, UVOT), CXO, HST, RATIR, LCOGT (FTN), Liverpool:2m, Magellan, Maidanak:1.5m, CrAO:1.25m, VLA, ATCA}

\begin{table*}
  \centering
 \caption{Model Comparison }
  \begin{tabular}{cccc}
    \toprule
& Gaussian (free $\xi$) & Gaussian (fixed $\xi$) & Top Hat\\
    
\mbox{WAIC} & 1782.7 $\pm$ 79.1 & 1744.3 $\pm$ 78.5 & -3561.8 $\pm$ 167.2    \\
    
$\Delta \mbox{WAIC} / \mbox{N}$ & - & 0.10 $\pm$ 0.09 & -14.5 $\pm$ 2.7  \\

Confidence Level & - & (0.58-1.16)$\sigma$ & (2.7-5.3)$\sigma$ \\
\hline
  \end{tabular}
   \begin{tablenotes}
      \small
      \centering
      \item A WAIC analysis is performed for each model considered. A higher WAIC score indicates better predictability for future data based on the model. The Gaussian model with $\xi$ free to vary has a higher likelihood of describing new data well so we use it as the base to compare the others. Each Gaussian model has a higher WAIC score compared to the top-hat case but there is no strong preference for the free-$\xi$ over the fixed-$\xi$ Gaussian model.
    \end{tablenotes}
  \label{tab:modelcomp}
\end{table*}

    
    


\bibliography{ref}{}
\bibliographystyle{aasjournal}





     

\end{document}